\documentclass[10pt,letterpaper,twocolumn]{article} 

\usepackage{ol2}
\usepackage[draft]{hyperref}
\usepackage{amsmath}

\begin{document}

\twocolumn[  

\title{Designing and understanding directional emission from spiral microlasers}

\author{Martina Hentschel and Tae-Yoon Kwon}
\address{Max-Planck-Institut f\"ur Physik komplexer Systeme,
  N\"othnitzer Str.~38, D-01187 Dresden, Germany}
\date{\today}

%%%%%%%%%%%%%%%%%%% abstract and OCIS codes %%%%%%%%%%%%%%%%
%% [use \begin{abstract*}...\end{abstract*} if exempt from copyright]

\begin{abstract}
The availability of microlasers with highly directional far-field characteristics is crucial for future applications. To this end we study the far-field emission of active microcavities with spiral shape using the Schr\"odinger-Bloch model. We find that they can provide directional emission under the conditions 
of (i) pumping along the resonator boundary and (ii) for specific resonator geometries. We systematically study the far-field characteristics under variation of the pumped area and the cavity geometry, and identify an directionality-optimized regime. Our results consistently explain previously obtained experimental results.
\end{abstract}


]

%%%%%%%%%%%%%%%%%%%%%%%%%%  body  %%%%%%%%%%%%%%%%%%%%%%%%%%

%-----------------------------------------------------------
%Introduction
%-----------------------------------------------------------
The quest to achieve directional emission from disk-like optical 
microcavities and microlasers began, literally, with the fabrication 
of the first microdisk lasers \cite{mccall}. A microdisk emits 
light uniformly in all directions due to its rotational invariance. 
Although evanescent emission ensures the, usually desired, high 
$Q$-factors, uniform emission 
is a serious drawback for practical applications such as light 
amplification by microlasers in integrated, fiber-optic based 
photonic devices and circuits\cite{chang,vahala}. 

One line of research followed over the past years concerns 
the potential applicability of spiral microcavities to achieve directional emission. 
Their boundary is defined, in polar coordinates ($r,\phi$), as $r(\phi)= R_0 (1 + \epsilon \phi/360^\circ)$ where $R_0$ gives the minimal radius at $\phi=0$ that has increased 
to $(1+\epsilon) R_0 $ when reaching $\phi=360^\circ$. 
The application potential of spiral cavities is seen, in short, in the symmetry breaking induced by 
the so-called notch (the straight line closing the boundary at $\phi=0=360^\circ$) 
that might lead to directional emission. 

Indeed, such a directional emission has been observed in various experiments
\cite{yale_spiral_optical,yale_spiral_electrical,french_spiral,japanese_spiral,korean_spiral_apl} under an angle of $\phi \approx 45^\circ$ with respect 
to the notch line.
A naive explanation, to be considerably refined as we shall see below,   
would be that whispering gallery modes traveling such that they
hit the notch are coupled out refractively in the observed direction. 
Diffraction effects of the whispering gallery modes not hitting the notch 
were discussed in \cite{yale_spiral_optical}, but no 
supporting arguments are found in the present study.

Very recent results on the radiation characteristics of quantum-cascade (QC) spiral 
microcavity lasers \cite{capasso_apl,hentschel_pra} report, however, 
the lack of a systematic
emission directionality, in contrast to the above-mentioned results. Note that the QC lasers are always operated in TM polarization and that they were uniformly pumped, unlike the other works where mostly boundary pumping was used. 
We also point out that it is well known that TE polarization leads to a better directionality \cite{gmachl,limacon,hentschel_pra}.
The non-directional emission 
found for QC microspirals \cite{capasso_apl} is in agreement with ray and wave simulations on Fresnel billiards 
of spiral shape \cite{hentschel_pra} and is an expression of the resonator's 
chaotic dynamics.

In the present letter, we address the question when, and why, 
directional emission from spiral-shaped microcavity lasers can be expected. 
To this end we simulate the far-field radiation characteristics of TM-polarized light in the framework of the Schr\"odinger-Bloch model \cite{schroedingerbloch} for different set-ups motivated by those used in the experiments. In particular, we vary the resonator geometry (parameter $\epsilon$) and the area of the continuously pumped region (boundary vs. uniform pumping schemes).

\begin{figure}[tb]
\begin{center}
  \includegraphics[width=7.8cm]{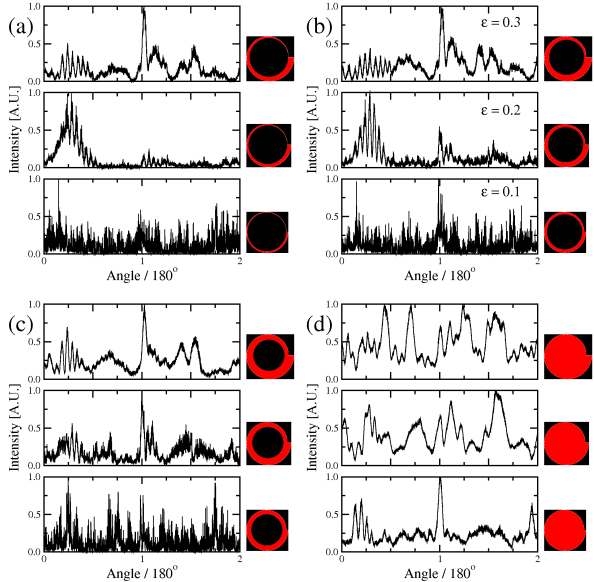}  %was 8.5cm
\end{center}  
  \caption{(Color online) Far-field radiation characteristics for spiral cavities of three different geometries ($\epsilon=$0.3, 0.2, 0.1 from top to bottom) and three different boundary pumping schemes, (a)-(c), as well as for uniform pumping, (d). The pumped areas are shown in red in the insets, the parameters are $R/R_0=$ 0.99, 0.9, 0.8 and 0 from (a) to (d). The intensity is obtained as the time average after a transient regime when stable lasing operation is established. 
  It is normalized to one in each panel. The size parameter (gain center) is $n k R_0 \approx$ 62.2, similar values yield comparable results. 
  }
  \label{fig2}
\end{figure}

Figure \ref{fig2} shows the far-field patterns (taken at a distance of $3R_0$ \cite{hentschel_pra}) for different geometries and pumping schemes. The far-field intensity is plotted against the far-field angle $\phi$.
Panels (a)-(d) correspond to different pumping schemes, and the pumped area is 
shown in red in each of the right insets. 
Pumping is performed everywhere except in a circle of radius $R<R_0$. 
For each of the pumping schemes $R/R_0=$ 0.99, 0.9, 0.8 and 0, three different 
geometries are considered, namely $\epsilon=$ 0.1, 0.2, and 0.3. 
A refractive index of $n=3.15$ was used and the parameters used in the Schr\"odinger-Bloch model are given in Ref.~\onlinecite{hentschel_pra}. %and correspond to QC lasers. 
A size parameter $n k R_0 \approx$ 62.2 (corresponding to the gain center) is used throughout the paper, here $k = 2 \pi/\lambda$ is the wavenumber with $\lambda$ being the wavelength in vacuum. For a cavity size of $R_0=22 \mu m$, we then find $\lambda=7 \mu m$. The size parameter is chosen such that the notch size $\epsilon R_0$ is approximately equal to $\lambda/n$ ($2 \lambda/n$, $3 \lambda/n$) for $\epsilon=0.1$ (0.2, 0.3).

The poorest directionality is observed for uniform pumping, panel (d). For the boundary pumping schemes, a pronounced dependence on both the geometry and the pumped area is observed. Note that directional emission (indicated by the existence of, ideally, only one peak in the far-field characteristics) occurs under a far-field angle of $\phi \approx 45^\circ$ and 
is related to emission from the notch as we shall see below. A rather sharp peak can occur at $\phi \approx 
180^\circ$; it is, however, always accompanied by a number of broad and pronounced side peaks.

\begin{figure}[tb]
\begin{center}
  \includegraphics[width=7.8cm]{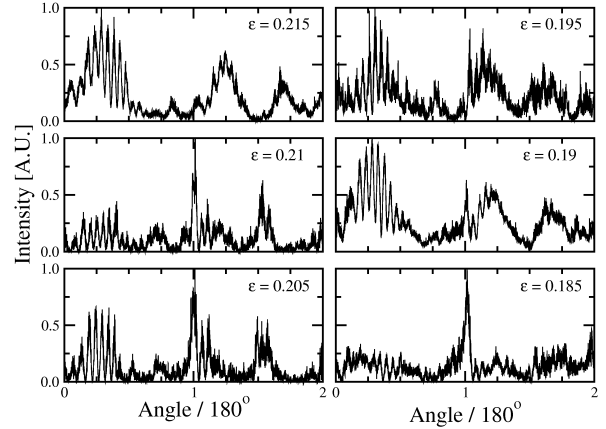}  %was 8.5cm
\end{center}
  \caption{Far-field characteristics for boundary-pumped spiral microcavities ($R/R_0=0.9$) as in Fig.~\ref{fig2}(b), but now for geometry parameters $\epsilon$ around 0.2. 
  The chaotic dynamics
  in the spiral billiards results in a strong sensitivity against variations of the geometry, causing shifts in the relative weight of the emission peaks. The directed emission from the notch under an angle $\phi \approx 45^\circ$ persists for most $\epsilon$ near 0.2.
  }
  \label{fig3}
\end{figure}

For all boundary-pumped schemes, cf.~Figs.~\ref{fig2}(a)-(c), the geometry $\epsilon=0.1$ is the least favorable with respect to directionality: The far-field output is almost uniform. As this shape is also the one closest to the disk (circular geometry), this finding illustrates clearly the importance of the symmetry breaking (deviation from the disk geometry, parametrized by the notch size $\epsilon$) in order to achieve directional light output. However, directed emission is lost likewise for too large $\epsilon$, see the panels for $\epsilon=0.3$, that are characterized by a number of mostly broad peaks. The optimal geometry seems to be in between, around $\epsilon \approx 0.2 \cong 2 \lambda / (n R_0)$, as we also confirm in Fig.~\ref{fig3}.  That is, the best directionality is found for neither too small nor too large notch size, and when pumping is performed very close to the resonator boundary, cf.~Figs.~\ref{fig2}(a) and (b). Whereas the pumping scheme in Fig.~\ref{fig2}(a), $R/R_0=0.99$, shows the best directionality, we will focus on $R/R_0=0.9$, Fig.~\ref{fig2}(b), in the following: Its experimental realization appears to be more realistic as (i) the pumped area may remain finite at the notch and (ii) %, related to the larger pumped area, 
a lower lasing threshold can be expected.
%as it does not require the pumped area to be restricted to almost zero next to the notch. 

In Fig.~\ref{fig3}, we show the evolution of the far-field pattern in the parameter range around $\epsilon \approx 0.2$. For most parameters the emission from the notch is clearly visible, accompanied by various other peaks of varying height. This, as well as the vanishing of emission generating from the notch at certain parameters, is due to the intrinsically chaotic light (ray) dynamics in spiral microcavities \cite{hentschel_pra} that results in a sensitive dependence on all parameters, here in particular the geometry parameter $\epsilon$.

\begin{figure}[tb]
  \centering
  \includegraphics[width=5.5cm]{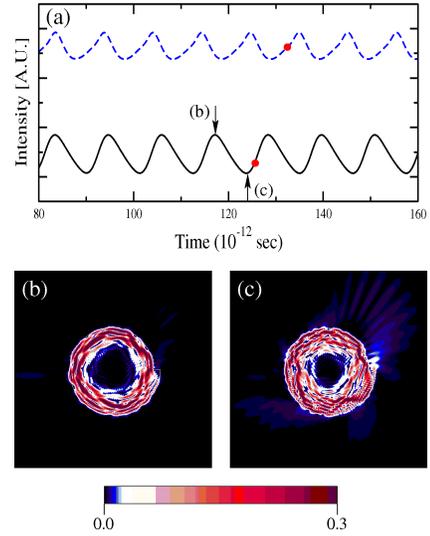}  %was 7cm
  \caption{(Color online) (a) Total internal light intensity in the lasing regime as a function of time for near-threshold external pumping strengths of $W_{\infty}=0.0002$ (solid line) and $W_{\infty}=0.0004$ (dashed line; this value is used in all other figures and in Ref.~\onlinecite{hentschel_pra}) for $R/R_0=0.9$ and $\epsilon=0.2$. The oscillations resemble those known from two-mode beating. 
Oscillations take place between a high-$Q$ cavity state, see snapshot (b), and a low-$Q$ cavity state, see snapshot (c). The directed output from the notch originates from cavity states similar to snapshot (c). The output is pulsed and highest at the red circles marked in (a). 
  In snapshots (b) and (c) the intensity is normalized to one, values larger than 0.3 are shown in the same color in order to highlight the cavity's emission properties.
 }
  \label{fig1}
\end{figure}

Finally, we turn to the question what mechanism generates the directed emission from the notch. Figure \ref{fig1}(a) shows the total light intensity inside the cavity for the boundary pumping scheme of Fig.~\ref{fig2}(b) and $\epsilon=0.2$, as a function of time in the lasing regime. 
Its sinusoidal oscillations are reminiscent of the beating between two modes observed in the intermediate-to-strong pumping regime of microlasers\cite{schroedingerbloch,harayama2}.
This picture is supported by the decrease of the period at increased pumping strength (dashed line), accompanied by a distortion of the sinusoidal oscillations. We checked that Fourier transformation of the %total 
intensity signal yields two dominant peaks;
their frequency difference gives the oscillation period. 
% and that their frequency difference determines the oscillation period. 

Strictly speaking, mode-beating assumes the existence of two (or more) resonances or stationary states. Those are well defined in uniformly pumped cavities \cite{schroedingerbloch}, when a one-to-one correspondence exists between resonances of the passive cavity and lasing modes. In the present case of boundary pumping, the definition of stationary states is not straightforward. Nonetheless, two cavity states, associated with maximal and minimal intensity inside the cavity and shown in Figs.~\ref{fig1}(b) and (c), can be distinguished and we will refer to them as high-$Q$ and low-$Q$ mode.  
They differ in both their emission properties and their intensity distribution inside the cavity. In the snapshot shown in Fig.~\ref{fig1}(b), the intensity is concentrated at the resonator boundary, whereas it penetrates further towards the cavity center in the snapshot in Fig.~\ref{fig1}(c). Cavity state (b) is closer to whispering-gallery or quasi-scarred \cite{quasiscar} states, it has the higher $Q$-factor ($Q \stackrel{>}{\small \sim} 1000$, say) and is less leaky. 

A classification of these two cavity states in terms of angular momenta is also useful and readily explains these differences: Snapshots (b) [(c)] are dominated by [counter-] clockwise-propagating components. The dominance of counterclockwise-propagating components
(note that those exist only in boundary-pumping schemes) increases for larger $R/R_0$ and improves the directional output from the notch, cf.~Figs.~\ref{fig2}(a) and (b). For too small $R/R_0$, directed output is lost, Fig.~\ref{fig2}(c.)

Note that two-mode beating typically occurs between modes of similar frequency and $Q$-factor\cite{schroedingerbloch}. The beating-type interaction between a high-$Q$ and a low-$Q$ state shown in Fig.~\ref{fig1}(a) contrasts this conventional dynamics. 
The interaction is the result of boundary pumping and would not be possible otherwise.
Boundary pumping allows both the high-$Q$ and the low-$Q$ state to experience a comparable gain (for certain $\epsilon$ and $R/R_0$ sufficiently close to one, cf.~Fig.~\ref{fig2}), and a beating-type interaction becomes possible.

The emission from the notch occurs in a pulsed fashion as the system oscillates between the two cavity states (b) and (c). 
It occurs predominately when the system is close to cavity state (c), 
i.e., towards and in minima of the total internal intensity. 
%Emission from the notch area increases towards the minima of the total internal intensity. 
The maximum emission from the notch is reached just after passing through the minimum and marked by red circles in Fig.~\ref{fig1}(a). 

{\it Conclusions.} We have shown and explained that directional emission from spiral microdisk lasers is possible with boundary-pumping schemes, but not with uniform pumping,
in full agreement with experimental observations. The directionality improves for smaller pumping areas $R/R_0 \rightarrow 1$, but values $R/R_0 \approx 0.9$ already lead to a pronounced output directionality ($\phi \approx 45^\circ$).
The cavity geometry parameter $\epsilon$ has also to be tuned and 
we have found that shapes that differ sufficiently, 
but not too much 
from a circular cavity provide the best directionality. We have found directional emission around $\epsilon \approx 0.2$, corresponding to a notch length of about twice the cavity wavelength $\lambda/n$. The directed output occurs in a pulsed manner from a low-$Q$ cavity state 
and %results 
from a mechanism that is reminiscent of two-mode beating.

{\it Acknowledgment.} M.H. thanks the German Research Foundation (DFG) for support in the
Emmy-Noether Program and the Research Group FG 760. T.-Y. Kwon was partly supported by the ``Korea Research Foundation Grant'' funded by the Korean government (MOEHRD) (contract number KRF-2006-352-C00022).

%%%%%%%%%%%%%%%%%%%%%%% References %%%%%%%%%%%%%%%%%%%%%%%%%

\end{document}